\documentclass{birkjour}
\usepackage{amssymb}
\usepackage{amsmath}

\def\diag{{\rm diag}}
\def\cl{{ C}\!\ell}
\def\Even{{\rm Even}}
\def\Odd{{\rm Odd}}
\def\Tr{{\rm Tr}}
\def\R{{\Bbb R}}
\def\C{{\Bbb C}}
\def\F{{\Bbb F}}
\def\H{{\Bbb H}}
\def\Mat{{\rm Mat}}
\def\SO{{\rm SO}}
\def\O{{\rm O}}
\def\Spin{{\rm Spin}}
\def\Pin{{\rm Pin}}
\newcommand{\st}{\stackrel}

\def\I{{I}}

 \newtheorem{thm}{Theorem}[section]

 \theoremstyle{definition}
 
 \theoremstyle{remark}

 \numberwithin{equation}{section}

\begin{document}

%
%
%
%
%
%
%
%
%

\title[Calculation of elements of spin groups]
 {Calculation of elements of spin groups using generalized Pauli's theorem}

\author[Dmitry Shirokov]{Dmitry Shirokov}

\address{%
A.A.Kharkevich Institute for Information Transmission Problems\\
Russian Academy of Sciences\\
Bol'shoi Karetnyi per., 19\\
127994, Moscow\\
Russia;\\
N.E.Bauman Moscow State University\\
ul. Baumanskaya 2-ya, 5\\
105005, Moscow\\
Russia;
}

\email{dm.shirokov@gmail.com}



\date{October 1, 2012}

\begin{abstract}
We formulate generalizations of Pauli's theorem on the cases of real and complex Clifford algebras of even and odd dimensions. We give analogues of these theorems in matrix formalism. Using these theorems we present an algorithm for computing elements of spin groups that correspond to elements of orthogonal groups as double cover.
\end{abstract}

\maketitle
\section{Introduction}
In 1936 Pauli published \cite{Pauli} his fundamental theorem for the Dirac gamma matrices $\gamma^a$, $a=1, 2, 3, 4$. He was interested in relation between 2 different sets of $\gamma$-matrices.

In this paper we formulate generalizations \cite{DAN2} of Pauli's theorem on the cases of real and complex Clifford algebras of even and odd dimensions. These theorems generalize well-known statement that there is one irreducible representation of Clifford algebra in even dimensions and there are two inequivalent irreducible  representations in odd dimensions. We consider a more general question about relations between two sets of Clifford algebra elements that satisfy anticommutation relations. It is shown that in real odd case there are 4 (6 in complex case) different cases of relation between two sets.

Using generalized Pauli's theorem we present an algorithm for computing elements of spin groups that correspond to elements of orthogonal groups as double cover. In our work we use formalism of Clifford algebras.

\section{Clifford algebras over the field of real and complex numbers}
There are several different (equivalent) definitions of Clifford algebras. For example, you can find 5 different definitions of Clifford algebra in \cite{Lounesto}, suitable for different purposes. The most popular is definition of Clifford algebra as a quotient of the tensor algebra when we have quadratic form on the vector space $V\subset \cl(p,q)$. In this paper we use another definition for our purpose. In our consideration it will be more convenient to use definition of Clifford algebra with the fixed bases (see \cite{MarchukShirokov}, \cite{Lounesto}) - enumerated by the ordered multi-indices. Note that generators and basis elements are fixed.

Let $E$ be a vector space over the field of real $\R$ or complex $\C$ numbers. Dimension of $E$ equals $2^n$, where $n$ is a natural number. Let we have a basis in $E$
\begin{equation}
e,\, e^a,\, e^{a_1 a_2},\,\ldots, e^{1\ldots n},\quad \mbox{where} \, a_1<a_2<\ldots,\qquad \mbox{($2^n$ elements)}\label{basis}
\end{equation}
enumerated by the ordered multi-indices of length from $0$ to $n$. Indices $a, a_1, a_2,\ldots$ take the values from $1$ to $n$.

Let $p$ and $q$ be nonnegative integer numbers such that $p+q=n$, $n\geq 1$. Consider the diagonal matrix
$$\eta=||\eta^{ab}||=\diag(1,\ldots,1,-1,\ldots,-1),$$
whose diagonal contains $p$ elements equal to $+1$ and $q$ elements equal to $-1$.

We introduce the operation of {\it Clifford multiplication} $U,V\to UV$ on $E$ such that we have the properties of distributivity, associativity, $e$ is identity element and
$$e^{a_1}\ldots e^{a_k}=e^{a_1\ldots a_k},\qquad 1\leq a_1<\ldots a_k\leq n,$$
$$e^a e^b+ e^b e^a=2\eta^{ab}e,\qquad \forall a,b=1,\ldots n.$$

Then introduced in this way algebra is called {\it real (complex) Clifford algebra} and it is denoted by $\cl^\R(p,q)$ or $\cl^\C(p,q)=\cl(p,q)$. When results are true for both cases, we write $\cl^\F(p,q)$, where $\F=\R$ or $\F=\C$.

Note, that many authors consider only complex Clifford algebras $\cl(n,0)=\cl(n)$ of the signature $(n,0)$ because $\cl(n,0)\simeq\cl(p,q)$ for $\forall p, q: p+q=n$ (see Theorem 3). But if we consider $\cl(n,0)$ and $\cl(p,q)$ as not just algebras, but as Clifford algebras, they are 2 different objects. For example, we have different action of operation of complex conjugation in these two Clifford algebras. That's why, when we consider Dirac equation we use $\cl(1,3)$, not $\cl(4,0)$.

Any Clifford algebra element $U\in\cl^\F(p,q)$ can be written in the form
\begin{eqnarray}
U=ue+u_a e^a+\sum_{a_1<a_2}u_{a_1 a_2}e^{a_1 a_2}+\ldots+u_{1\ldots n}e^{1\ldots n},\label{U}
\end{eqnarray}
where $u, u_a, u_{a_1 a_2}, \ldots, u_{1\ldots n}$ are real (complex) numbers.
We denote the vector subspaces spanned by the elements
$e^{a_1\ldots a_k}$ enumerated by the ordered multi-indices of length $k$ by $\cl^\F_k(p,q)$. The elements of the subspace $\cl^\F_k(p,q)$ are called {\it elements of rank $k$}. We have
\begin{eqnarray}
\cl^\F(p,q)=\bigoplus_{k=0}^{n}\cl^\F_k(p,q).\label{bigop}
\end{eqnarray}

Clifford algebra $\cl^\F(p,q)$ is a superalgebra. It is represented as the direct sum of even and odd subspaces
$$\cl^\F(p,q)=\cl^\F_{\Even}(p,q)\oplus\cl^\F_{\Odd}(p,q),$$ where
$$\cl^\F_{\Even}(p,q)=\bigoplus_{k - even}\cl^\F_k(p,q),\qquad \cl^\F_{\Odd}(p,q)=\bigoplus_{k - odd}\cl^\F_k(p,q).$$

Suppose, $U\in\cl^\F(p,q)$ is written in the form (\ref{U}).
Then denote $$\langle U\rangle_k=\st{k}{U}=\sum_{a_1<\cdots<a_k}u_{a_1\ldots
a_k}e^{a_1\ldots a_k}\in\cl_k(p,q).$$

Using the projection operator to the $1$-dimensional vector space
$\cl^\F_0(p,q)$ we define the
{\it trace} of an element $U\in\cl^\F(p,q)$ as
$$\Tr(U)=\langle U\rangle_0|_{e\to 1}=u.$$

The main property of the trace is
$$\Tr(UV)=\Tr(VU).$$

Consider the following operations of {\it grade involution} and {\it reversion}\footnote{We use notations from \cite{Lounesto}.} in $\cl^\F(p,q)$:
\begin{eqnarray}
U^{\wedge}=U|_{e^a\to-e^a},\quad
U^\sim=U|_{e^{a_1\ldots a_r}\to e^{a_r}\ldots e^{a_1}}.\label{wedge}
\end{eqnarray}
We have
$$U^{\wedge\wedge}=U,\quad U^{\sim\sim}=U,\quad (UV)^{\wedge}=U^{\wedge} V^{\wedge},\quad (UV)^\sim=V^\sim U^\sim$$
and
\begin{eqnarray*}
U^{\wedge}&=\sum_{k=0}^n (-1)^k\langle U\rangle_k&=
\langle U\rangle_0-\langle U\rangle_1+\langle U\rangle_2-
\langle U\rangle_3+\langle U\rangle_4-\ldots,\\
U^\sim&=\sum_{k=0}^n (-1)^{\frac{k(k-1)}{2}} \langle U\rangle_k&=
\langle U\rangle_0+\langle U\rangle_1-\langle U\rangle_2-
\langle U\rangle_3+\langle U\rangle_4+\ldots
\end{eqnarray*}

We have the following well-known statement about the center
$$cen\cl^\F(p,q)=\{U\in \cl(p,q) \,|\, UV=VU\quad \forall V\in\cl^\F(p,q)\}$$ of Clifford algebra.

\begin{thm}\label{theoremCentr} The center $cen\cl^\F(p,q)$ of Clifford algebra $\cl^\F(p,q)$ of dimension $n=p+q$ is the following subspace:
\begin{equation}
cen\cl^\F(p,q)=\left\lbrace
\begin{array}{ll}
\cl^\F_0(p,q), & \parbox{.5\linewidth}{ in the case of odd $n$;}\\
\cl^\F_0(p,q)\oplus\cl^\F_n(p,q), & \parbox{.5\linewidth}{ in the case of even $n$.}
\end{array}\nonumber
\right.
\end{equation}
\end{thm}

All real and complex Clifford algebras are isomorphic as algebras to matrix algebras. We have the following well-known theorems.
\begin{thm} (Cartan, Bott) \label{theoremReal}We have the following algebra isomorphisms for real Clifford algebra:
\begin{equation}
\cl^\R(p,q)\simeq\left\lbrace
\begin{array}{ll}
\Mat(2^{\frac{n}{2}},\R), & \parbox{.5\linewidth}{ if $p-q\equiv0; 2\!\!\mod 8$;}\\
\Mat(2^{\frac{n-1}{2}},\R)\oplus \Mat(2^{\frac{n-1}{2}},\R), & \parbox{.5\linewidth}{ if $p-q\equiv1\!\!\mod 8$;}\\
\Mat(2^{\frac{n-1}{2}},\C), & \parbox{.5\linewidth}{ if $p-q\equiv3; 7\!\!\mod 8$;}\\
\Mat(2^{\frac{n-2}{2}},\H), & \parbox{.5\linewidth}{ if $p-q\equiv4; 6\!\!\mod 8$;}\\
\Mat(2^{\frac{n-3}{2}},\H)\oplus \Mat(2^{\frac{n-3}{2}},\H), & \parbox{.5\linewidth}{ if $p-q\equiv5\!\!\mod 8$.}
\end{array}\nonumber
\right.
\end{equation}
\end{thm}

\begin{thm}\label{theoremCompl}We have the following algebra isomorphisms for complex Clifford algebra:
\begin{eqnarray}
\cl(p,q)\simeq\left\lbrace
\begin{array}{ll}
\Mat(2^{\frac{n}{2}}, \C), & \parbox{.5\linewidth}{if $n$ - even;}\\
\Mat(2^{\frac{n-1}{2}}, \C)\oplus \Mat(2^{\frac{n-1}{2}}, \C), & \parbox{.5\linewidth}{if $n$ - odd.}
\end{array}\nonumber
\right.
\end{eqnarray}
\end{thm}

\section{Generalized Pauli's theorem in Clifford algebras}

In 1936 Pauli published \cite{Pauli} his fundamental theorem for Dirac gamma-matrices $\gamma^a$, $a=1, 2, 3, 4$. He was interested in relation between 2 different sets of $\gamma$-matrices.

\begin{thm}[Pauli, \cite{Pauli}] Let two sets of square complex matrices $$\gamma^a,\qquad \beta^a,\qquad a=1, 2, 3, 4$$
of order $4$ satisfy the relations
\begin{eqnarray}
\gamma^a \gamma^b + \gamma^b \gamma^a&=& 2 \eta^{ab} {\bf1}, \label{satisf}\\
\beta^a \beta^b + \beta^b \beta^a&=& 2 \eta^{ab} {\bf1},\nonumber
\end{eqnarray}
where $\eta=|\eta^{ab}|=\diag(1, -1, -1, -1)$ is a diagonal matrix.

Then there exists a unique (up to a multiplicative nonzero complex constant) invertible matrix $T$ such that
$$\gamma^{a}=T^{-1}\beta^a T,\qquad a=1, 2, 3, 4.$$
\end{thm}

In \cite{DAN2} we present generalizations of this theorem. Let formulate these generalizations for the cases of real and complex Clifford algebras of even and odd dimensions.

Let denote multi-index of arbitrary length by $A$ and denote its length by $|A|$. Expression $\gamma^{a_1} \ldots \gamma^{a_k}$ is denoted by $\gamma^{a_1\ldots a_k}$ for $a_1<\ldots<a_k$.
We have the following notations for sets of multi-indices:
\begin{eqnarray}
\I&=&\{\o, 1, \ldots, n, 12, 13, \ldots, 1\ldots n\},\nonumber\\
\I_{\Even}&=&\{A\in\I,\quad |A| - \mbox{even}\},\nonumber\\
\I_{\Odd}&=&\{A\in\I,\quad |A| - \mbox{odd}\}.
\end{eqnarray}

At first, let formulate generalization of Pauli's theorem for the case of arbitrary even dimension $n$.

\begin{thm}\label{theoremPauliEv} Consider real (or, respectively complex) Clifford algebra $\cl^\F(p,q)$ of even dimension $n=p+q$. Let two sets of Clifford algebra elements
\begin{eqnarray}\gamma^a,\qquad \beta^a,\qquad a=1, 2, \ldots, n\nonumber\end{eqnarray}
satisfy conditions $$\gamma^a \gamma^b + \gamma^b \gamma^a= 2 \eta^{ab} e,$$ $$\beta^a \beta^b + \beta^b \beta^a= 2 \eta^{ab} e.$$

Then both sets generate bases of Clifford algebra and there exists a unique (up to multiplication by a nonzero real (respectively, complex) constant) invertible element $T\in\cl^\F(p,q)$ such that
\begin{eqnarray}
\gamma^{a}=T^{-1}\beta^a T,\qquad \forall a=1, \ldots, n. \nonumber
\end{eqnarray}
Moreover, $T$ can be written in the following form
$$T=\beta^A F \gamma_A,$$
where $F$ is such element from the set
\begin{itemize}
  \item $\{ \gamma^A, A\in\I_{\Even}\}$ if $\beta^{1\ldots n}\neq-\gamma^{1\ldots n}$,
  \item $\{ \gamma^A, A\in\I_{\Odd}\}$ if $\beta^{1\ldots n}\neq\gamma^{1\ldots n}$,
\end{itemize}
that $\beta^A F \gamma_A\neq 0$.
\end{thm}

Now let formulate theorems for real and complex Clifford algebras of odd dimension $n$.

\begin{thm}\label{theoremPauliOddReal} Consider real Clifford algebra $\cl^\R(p,q)$ of odd dimension $n=p+q$. Suppose that 2 sets of Clifford algebra elements
\begin{eqnarray}\gamma^a,\, \beta^a,\qquad a=1, 2, \ldots, n\nonumber\end{eqnarray}
satisfy the relations $$\gamma^a \gamma^b + \gamma^b \gamma^a= 2 \eta^{ab} e,$$ $$\beta^a \beta^b + \beta^b \beta^a= 2 \eta^{ab} e.$$

Then, in Clifford algebra $\cl^\R(p,q)$ of signature $p-q\equiv1\!\!\mod4$ the elements $\gamma^{1\ldots n}$ and $\beta^{1\ldots n}$ take the values $\pm e^{1\ldots n}$ if and only if the corresponding sets generate bases of Clifford algebra. They take the values $\pm e$ if and only if the sets do not generate bases. In this situation, we have cases 1-4 below.

In Clifford algebra $\cl^\R(p,q)$ of signature $p-q\equiv3\!\!\mod4$ the elements $\gamma^{1\ldots n}$ and $\beta^{1\ldots n}$ always take the values $\pm e^{1\ldots n}$ and the corresponding sets always generate bases of the Clifford algebra. In this situation, cases 1 and 2 only hold.

There exists a unique (up to multiplication by an invertible element of the center of the Clifford algebra) invertible element $T$ of the Clifford algebra such that
\begin{eqnarray}
&1.& \gamma^{a}=T^{-1}\beta^a T,\quad  \forall a=1, \ldots, n  \quad\Leftrightarrow\, \beta^{1\ldots n}=\gamma^{1\ldots n}\nonumber;\\
&2.& \gamma^{a}=-T^{-1}\beta^a T,\quad  \forall a=1, \ldots, n  \quad\Leftrightarrow\, \beta^{1\ldots n}=-\gamma^{1\ldots n} \nonumber;\\
&3.& \gamma^{a}=e^{1\ldots n}T^{-1}\beta^a T,\quad  \forall a=1, \ldots, n  \quad\Leftrightarrow\, \beta^{1\ldots n}=e^{1\ldots n}\gamma^{1\ldots n}\nonumber;\\
&4.& \gamma^{a}=-e^{1\ldots n}T^{-1}\beta^a T,\quad  \forall a=1, \ldots, n  \quad\Leftrightarrow\, \beta^{1\ldots n}=-e^{1\ldots n}\gamma^{1\ldots n}. \nonumber
\end{eqnarray}

Note that all 4 cases have the unified notation $$\gamma^a=(\beta^{1\ldots n}\gamma_{1\ldots n})T^{-1}\beta^a T.$$

Additionally, in the case of real Clifford algebra of signature $p-q\equiv1\!\!\mod 4$, the element $T$, whose existence is stated in all 4 cases of the theorem, can be written in the following form
\begin{eqnarray}T=\sum_{A\in\I_{\Even}}\beta^A F \gamma_A,\label{form33}\end{eqnarray}
where $F$ is an element from the set $\{ \gamma^A+\gamma^B,\, A, B\in\I_{\Even}\}.$

In the case of real Clifford algebra of signature $p-q\equiv3\!\!\mod4$, the element $T$, whose existence is stated in 1-2 cases of the theorem, can be written in the form (\ref{form33}), where $F$ such element from the set $\{\gamma^A,\, A\in\I_{\Even}\}$ that element (\ref{form33}) is nonzero.
\end{thm}

\begin{thm}\label{theoremPauliOddCompl} Consider complex Clifford algebra $\cl(p,q)$ of odd dimension $n=p+q$. Suppose that 2 sets of Clifford algebra elements
\begin{eqnarray}\gamma^a,\, \beta^a,\qquad a=1, 2, \ldots, n\nonumber\end{eqnarray}
satisfy the relations $$\gamma^a \gamma^b + \gamma^b \gamma^a= 2 \eta^{ab} e,$$ $$\beta^a \beta^b + \beta^b \beta^a= 2 \eta^{ab} e.$$

Then, in Clifford algebra $\cl(p,q)$ of the signature $p-q\equiv1\!\!\mod4$ the elements $\gamma^{1\ldots n}$ and $\beta^{1\ldots n}$ take the values $\pm e^{1\ldots n}$ if and only if the corresponding sets generate bases of the Clifford algebra. They take the values $\pm e$ if and only if the sets do not generate bases. In this situation, we have cases 1-4 below.

In Clifford algebra $\cl(p,q)$ of signature $p-q\equiv3\!\!\mod4$ the elements $\gamma^{1\ldots n}$ and $\beta^{1\ldots n}$ take the values $\pm e^{1\ldots n}$ if and only if the corresponding sets generate bases of the Clifford algebra. They take the values $\pm ie$ if and only if the sets do not generate bases. In this situation, we have cases 1, 2, 5 and 6 below.

There exists a unique (up to multiplication by an invertible element of the center of the Clifford algebra) invertible element $T$ of the Clifford algebra such that
\begin{eqnarray}
&1.& \gamma^{a}=T^{-1}\beta^a T,\quad  \forall a=1, \ldots, n  \quad\Leftrightarrow\, \beta^{1\ldots n}=\gamma^{1\ldots n}\nonumber;\\
&2.& \gamma^{a}=-T^{-1}\beta^a T,\quad  \forall a=1, \ldots, n  \quad\Leftrightarrow\, \beta^{1\ldots n}=-\gamma^{1\ldots n} \nonumber;\\
&3.& \gamma^{a}=e^{1\ldots n}T^{-1}\beta^a T,\quad  \forall a=1, \ldots, n  \quad\Leftrightarrow\, \beta^{1\ldots n}=e^{1\ldots n}\gamma^{1\ldots n}\nonumber;\\
&4.& \gamma^{a}=-e^{1\ldots n}T^{-1}\beta^a T,\quad  \forall a=1, \ldots, n  \quad\Leftrightarrow\, \beta^{1\ldots n}=-e^{1\ldots n}\gamma^{1\ldots n} \nonumber;\\
&5.& \gamma^{a}=ie^{1\ldots n}T^{-1}\beta^a T,\quad  \forall a=1, \ldots, n  \quad\Leftrightarrow\, \beta^{1\ldots n}=ie^{1\ldots n}\gamma^{1\ldots n}\nonumber;\\
&6.& \gamma^{a}=-ie^{1\ldots n}T^{-1}\beta^a T,\quad  \forall a=1, \ldots, n  \quad\Leftrightarrow\, \beta^{1\ldots n}=-ie^{1\ldots n}\gamma^{1\ldots n} \nonumber.
\end{eqnarray}

Note that all 6 cases have the unified notation $$\gamma^a=(\beta^{1\ldots n}\gamma_{1\ldots n})T^{-1}\beta^a T.$$

Additionally, the element $T$, whose existence is stated in all 6 cases of the theorem, can be written in the form (\ref{form33}) where $F$ is an element from the set $\{ \gamma^A+\gamma^B,\, A, B\in\I_{\Even}\}.$
\end{thm}

\section{Generalized Pauli's theorem in matrix formalism}
\label{section950sh}

Theorems \ref{theoremPauliEv}, \ref{theoremPauliOddReal} and \ref{theoremPauliOddCompl} can be reformulated in a matrix formalism using Theorems \ref{theoremReal} and \ref{theoremCompl}. We obtain the following theorems.

\begin{thm}\label{theoremPauliMatrEv} Let $n$ be a natural even number. Consider 2 sets of square matrices (of the same order)
\begin{eqnarray}
\gamma^a,\qquad \beta^a,\qquad a=1, 2, \ldots, n,
\end{eqnarray}
that satisfy the following conditions
\begin{eqnarray}
\gamma^a \gamma^b + \gamma^b \gamma^a&=& 2 \eta^{ab} {\bf1},\\
\beta^a \beta^b + \beta^b \beta^a&=& 2 \eta^{ab} {\bf1},\nonumber
\end{eqnarray}
where $\eta$ is a diagonal matrix of order $n$
\begin{equation}
\eta=||\eta^{ab}||=\diag(1,\ldots,1,-1,\ldots,-1),
\end{equation}
which diagonal contains $p$ elements equal to $+1$ and $q$ elements equal to $-1$.

We have the following statements.
\begin{itemize}
  \item Let these matrices are complex of order $2^{\frac{n}{2}}$. Then there exists a unique (up to multiplication by a nonzero complex constant) invertible matrix $T$ such that
      \begin{eqnarray}
      \gamma^{a}=T^{-1}\beta^a T,\qquad a=1, \ldots n.
      \end{eqnarray}
  \item Let the signature is $p-q\equiv0,2\!\!\mod 8$ and matrices are real of order $2^{\frac{n}{2}}$. Then there exists a unique (up to multiplication by a nonzero real constant) invertible matrix $T$ such that
      \begin{eqnarray}
      \gamma^{a}=T^{-1}\beta^a T,\qquad a=1, \ldots n.
      \end{eqnarray}
  \item Let the signature is $p-q\equiv4,6\!\!\mod8$ and matrices are over the quaternions of order $2^{\frac{n-2}{2}}$. Then there exists a unique (up to multiplication by a nonzero real constant) invertible matrix $T$ such that
      \begin{eqnarray}
      \gamma^{a}=T^{-1}\beta^a T,\qquad a=1, \ldots n.
      \end{eqnarray}
\end{itemize}
Moreover, matrix $T$ can be written in the following form $$T=\beta^A F\gamma_A,$$ where $F$ is such matrix from $\{\gamma^A, A\in\I_{\Even}\}$ that $\beta^A F\gamma_A\neq 0$.
\end{thm}

Note, that we have not restriction on the signature $(p,q)$ in the first case of theorem because this case corresponds to complex Clifford algebra $\cl^\C(p,q)$. We have a restriction on the signature $(p,q)$ in the second and the third cases of the theorem because these cases corresponds to real Clifford algebra $\cl^\R(p,q)$ (see Theorems \ref{theoremReal} and \ref{theoremCompl}). For example, if we consider matrix algebra $\Mat(2^{\frac{n}{2}},\R)$ in the case of signature $p-q\equiv6\!\!\mod 8$, $n=2$, $p=0$, $q=2$,
then there are no\footnote{Because real Clifford algebra corresponds to quaternion matrix algebras (not real matrix algebras) in the case of these signatures. We can obtain only quaternion representation.} matrices $\gamma^1, \gamma^2\in \Mat(2,\R)$ that
\begin{eqnarray}
(\gamma^1)^2=(\gamma^2)^2=-{\bf1},\quad \gamma^1 \gamma^2=-\gamma^2 \gamma^1.\label{u1}
\end{eqnarray}

Really, let we have these 2 matrices:
$$
\gamma^1=\left( \begin{array}{ll}
 a_1 & b_1 \\
 c_1 & d_1 \end{array}\right),\qquad
\gamma^2=\left( \begin{array}{ll}
 a_2 & b_2 \\
 c_2 & d_2 \end{array}\right),
$$
where $a_i, b_i, c_i, d_i \in\R, i=1, 2$. If we have (\ref{u1}),then
$$a_i=-d_i,\quad a_i^2+b_ic_i=-1,\quad i=1, 2,\quad 2a_1a_2+b_1c_2+b_2c_1=0.$$
We obtain $c_1\neq0$ and $c_2\neq 0$, so
$$2a_1a_2-\frac{1+a_1^2}{c_1}c_2-\frac{1+a_2^2}{c_2}c_1=0\quad \Rightarrow \quad-c_2^2-c_1^2=(c_2a_1-c_1a_2)^2.$$
Then $c_1=c_2=0$, and we have a contradiction.

Let formulate theorems for 2 sets of odd number of matrices.

We denote by $$J=\diag(1, \ldots, 1, -1, \ldots, -1)$$ the diagonal matrix of the required order, which diagonal contains the same number of elements equal to $1$ and $-1$.

\begin{thm}\label{theoremPauliMatrOdd} Let $n$ be a natural odd number. Consider 2 sets of square matrices (of the same order)
\begin{eqnarray}
\gamma^a,\qquad \beta^a,\qquad a=1, 2, \ldots, n,
\end{eqnarray}
that satisfy the following conditions
\begin{eqnarray}
\gamma^a \gamma^b + \gamma^b \gamma^a&=& 2 \eta^{ab} {\bf1},\\
\beta^a \beta^b + \beta^b \beta^a&=& 2 \eta^{ab} {\bf1},\nonumber
\end{eqnarray}
where $\eta$ is a diagonal matrix of order $n$
\begin{equation}
\eta=||\eta^{ab}||=\diag(1,\ldots,1,-1,\ldots,-1),
\end{equation}
which diagonal contains $p$ elements equal to $+1$ and $q$ elements equal to $-1$.

We have the following statements.
\begin{itemize}
  \item Let these matrices are complex\footnote{There are no restrictions on the signatures because this case corresponds to the complex (not real) Clifford algebra $\cl^\C(p,q)\simeq\Mat(2^{\frac{n-1}{2}}, \C)\oplus \Mat(2^{\frac{n-1}{2}}, \C)$, $n=p+q$ is odd.} on a diagonal.

  Then matrices $\beta^{1\ldots n}$ and $\gamma^{1\ldots n}$ equal to $\pm J$, $\pm iJ$ if and only if sets of matrices generate bases in $\Mat(2^{\frac{n-1}{2}},\C)\oplus \Mat(2^{\frac{n-1}{2}},\C)$. They equal to $\pm {\bf 1}$, $\pm i{\bf 1}$ if and only if sets of matrices don't generate bases.

  Then there exists an invertible matrix $T$ such that
      \begin{eqnarray}
      \gamma^{a}=\beta^{1\ldots n}\gamma_{1\ldots n}T^{-1}\beta^a T,\qquad a=1, \ldots n,
      \end{eqnarray}
      where $\beta^{1\ldots n}\gamma_{1\ldots n}=$ $\pm {\bf 1}$, $\pm i{\bf 1}$, $\pm J$, $\pm iJ$.

      Such matrix $T$ is unique up to multiplication by an invertible matrix of the form
      $\lambda {\bf1}+\mu J$, where $\lambda, \mu \in\C$.
  \item Let the signature is $p-q\equiv1\!\!\mod 8$ and matrices are real, block-diagonal of order $2^{\frac{n+1}{2}}$ with 2 blocks of order $2^{\frac{n-1}{2}}$ on a diagonal.

      Then matrices $\beta^{1\ldots n}$ and  $\gamma^{1\ldots n}$ equal to $\pm J$ if and only if sets of matrices generate bases in $\Mat(2^{\frac{n-1}{2}},\R)\oplus \Mat(2^{\frac{n-1}{2}},\R)$. They equal to $\pm {\bf 1}$ if and only if sets of matrices don't generate bases.

      Then there exists an invertible matrix $T$ such that
      \begin{eqnarray}
      \gamma^{a}=\beta^{1\ldots n}\gamma_{1\ldots n}T^{-1}\beta^a T,\qquad a=1, \ldots n,
      \end{eqnarray}
      where $\beta^{1\ldots n}\gamma_{1\ldots n}=\pm {\bf1}, \pm J$.

      Such matrix $T$ is unique up to multiplication by an invertible matrix of the form
      $\lambda {\bf1}+\mu J$, where $\lambda, \mu \in\R$.
\item Let the signature is $p-q\equiv5\!\!\mod 8$ and matrices are over the quaternions, block-diagonal of the order $2^{\frac{n-1}{2}}$ with 2 blocks of the order $2^{\frac{n-3}{2}}$ on a diagonal.

    Then matrices $\beta^{1\ldots n}$  and $\gamma^{1\ldots n}$ equal to $\pm J$ if and only if sets of matrices generate bases in $\Mat(2^{\frac{n-3}{2}},\H)\oplus \Mat(2^{\frac{n-3}{2}},\H)$. They equal to $\pm {\bf 1}$ if and only if sets of matrices don't generate bases.

    Then there exists an invertible matrix $T$ such that
      \begin{eqnarray}
      \gamma^{a}=\beta^{1\ldots n}\gamma_{1\ldots n}T^{-1}\beta^a T,\qquad a=1, \ldots n,
      \end{eqnarray}
      where $\beta^{1\ldots n}\gamma_{1\ldots n}=$ $\pm {\bf 1}$, $\pm J$.

      Such matrix $T$ is unique up to multiplication by an invertible matrix of the form
      $\lambda {\bf1}+\mu J$, where $\lambda, \mu \in\R$.

  \item Let the signature is $p-q\equiv3\!\!\mod 4$ and matrices are complex of order $2^{\frac{n-1}{2}}$. Then corresponding sets of matrices always generate bases.

      There exists an invertible matrix $T$ such that
      \begin{eqnarray}
      \gamma^{a}=\beta^{1\ldots n}\gamma_{1\ldots n}T^{-1}\beta^a T,\qquad a=1, \ldots n,
      \end{eqnarray}
      where $\beta^{1\ldots n}\gamma_{1\ldots n}=\pm {\bf1}$.
      Such matrix $T$ is unique up to multiplication by an invertible matrix of the form
      $\lambda {\bf1}+i \mu  J$, where $\lambda, \mu \in\R$.
\end{itemize}
In the 1), 2) and 3) cases of the theorem matrix $T$ can be written in the following form $$T=\sum_{A\in\I_{\Even}}\beta^A F\gamma_A,$$ where $F$ is an element from the set $\{\gamma^A+\gamma^B, A, B \in\I_{\Even}\}$.

In the 4) case of the theorem a matrix $T$ can be written in the form $$T=\sum_{A\in\I_{\Even}}\beta^A F\gamma_A,$$ where $F$ is such matrix from the set $\{\gamma^A, A\in\I_{\Even}\}$ that $\sum_{A\in\I_{\Even}}\beta^A F\gamma_A\neq 0$.
\end{thm}

Note, that we can choose such invertible matrix $T'$, that $T'^{-1}\gamma^aT'$ transforms block-diagonal matrices $\gamma^a$ in matrices that are not block-diagonal. For these sets of matrices Theorem \ref{theoremPauliMatrOdd} is also true.

Also note, that in the case of odd $n$ we can consider two times smaller matrices (that are not block-diagonal). They are not bases in corresponding matrix algebras (because they are linearly dependent), but they give an irreducible representation of Clifford algebra.

For example, we have the following statement. Let in the assumptions of the Theorem \ref{theoremPauliMatrOdd} matrices are complex of order $2^{\frac{n-1}{2}}$. Then there exists invertible matrix $T$ such that
      \begin{eqnarray}
      \gamma^{a}=\pm T^{-1}\beta^a T,\qquad a=1, \ldots n,
      \end{eqnarray}

Also we can generalize statements of this section on the matrices of larger order using Kronecker product.

Let give example that illustrate 4) case of Theorem \ref{theoremPauliMatrOdd}.

Let $p=3$, $q=0$. So $n=p+q=3$, $p-q\equiv3\!\!\mod4$. Consider the following set $\beta^a=\sigma^a$ of Pauli matrices
$$
\sigma^1=\left( \begin{array}{ll}
 0 & 1 \\
 1 & 0 \end{array}\right),\quad
\sigma^2=\left( \begin{array}{ll}
 0 & -i \\
 i & 0 \end{array}\right),\quad
\sigma^3=\left( \begin{array}{ll}
 1 & 0 \\
 0 & -1 \end{array}\right).
$$
and the set $\gamma^a=-\sigma^a$. Then $\beta^1 \beta^3 \beta^3=- \gamma^1 \gamma^2 \gamma^3$.

There is no matrix $T\in\Mat(2,\C)$ such that $\gamma^a=T^{-1}\beta^a T$.
Really, if we have such matrix $T$, then
$$\gamma^1 \gamma^2 \gamma^3=T^{-1}\beta^1 T T^{-1}\beta^2 T T^{-1} \beta^3 T=T^{-1} \beta^1 \beta^3 \beta^3 T =\beta^1 \beta^3 \beta^3,$$
(because $\beta^1 \beta^3 \beta^3=\sigma^1 \sigma^2 \sigma^3 = i {\bf1}$ commutes with all matrices) and we have a contradiction.

But there exists matrix $T=-{\bf1}$ such that $-\sigma^a=\gamma^a=T^{-1}\beta^a T$.

In the next sections of this paper we use formalism of Clifford algebras. But you can also use matrix formalism to calculate elements of spin groups and reformulate theorems from the next sections in matrix formalism.

\section{Generalized Pauli's theorem in the case of odd elements of Clifford algebra}

We can obtain the following corollaries from Theorems \ref{theoremPauliEv}, \ref{theoremPauliOddReal} and \ref{theoremPauliOddCompl}. We consider the case when $\gamma^a$ and $\beta^a$ are odd elements of Clifford algebra.

\begin{thm}\label{theoremPauliEvOdd} Consider real (or, respectively, complex) Clifford algebra $\cl^\F(p,q)$ of even dimension $n=p+q$. Let two sets of odd Clifford algebra elements
\begin{eqnarray}\gamma^a,\, \beta^a \in \cl^\F_{\Odd}(p,q),\qquad a=1, 2, \ldots, n\label{gmatr234}\end{eqnarray}
satisfy conditions
\begin{eqnarray}
\gamma^a \gamma^b + \gamma^b \gamma^a&=& 2 \eta^{ab} e, \label{gsootn25}\\
\beta^a \beta^b + \beta^b \beta^a&=& 2 \eta^{ab} e.\nonumber
\end{eqnarray}
Then both sets (\ref{gmatr234}) generate bases of Clifford algebra and $\gamma^{1\ldots n}$, $\beta^{1\ldots n}$ equal to $\pm e^{1\ldots n}$.

Moreover, there exists a unique (up to multiplication by a nonzero real (or, respectively, complex) constant) invertible element $T$ such that
\begin{eqnarray}
\gamma^{a}=T^{-1}\beta^a T,\qquad \forall a=1, \ldots, n.\label{pauli25}
\end{eqnarray}
Moreover,
\begin{itemize}
          \item $T\in\cl^\F_{\Even}(p,q)$ if $\beta^{1\ldots n}=\gamma^{1\ldots n}$,
          \item $T\in\cl^\F_{\Odd}(p,q)$ if $\beta^{1\ldots n}=-\gamma^{1\ldots n}$.
\end{itemize}
Moreover, such $T$ can be written in the following form
$$T=\beta^A F \gamma_A,$$
where $F$ is such element from the set
\begin{itemize}
  \item $\{ \gamma^A, A\in\I_{\Even}\}$ if $\beta^{1\ldots n}=\gamma^{1\ldots n}$,
  \item $\{ \gamma^A, A\in\I_{\Odd}\}$ if $\beta^{1\ldots n}=-\gamma^{1\ldots n}$,
\end{itemize}
that $T$ constructed on its $F$ is nonzero.
\end{thm}

\proof Note, that if $\gamma^a\in\cl^\F_{\Odd}(p,q)$, then $\gamma^{1\ldots n}=\pm e^{1\ldots n}$. Actually, from anticommutative conditions we obtain that $\gamma^{1\ldots n}$ commute with all even elements and anticommute with all odd elements. So, $\gamma^{1\ldots n}=\lambda e^{1\ldots n}$, $\lambda\in\F$. From $(e^{1\ldots n})^2=(\gamma^{1\ldots n})^2$ we obtain $\gamma^{1\ldots n}=\pm e^{1\ldots n}$.

So, we have 2 different cases: $\beta^{1\ldots n}=\gamma^{1\ldots n}$ and $\beta^{1\ldots n}=-\gamma^{1\ldots n}$. Then we use an algorithm for calculaing element $T$ from Theorem \ref{theoremPauliEv}. $\blacksquare$

\begin{thm}\label{theoremPauliOddOdd} Consider real (or, respectively, complex) Clifford algebra $\cl^\F(p,q)$ of odd dimension $n=p+q$. Let two sets of odd Clifford algebra elements
\begin{eqnarray}\gamma^a,\, \beta^a\in\cl^\F_{\Odd}(p,q),\qquad a=1, 2, \ldots, n\label{gmatr3}\end{eqnarray}
satisfy conditions
\begin{eqnarray}
\gamma^a \gamma^b + \gamma^b \gamma^a&=& 2 \eta^{ab} e, \label{gsootn3}\\
\beta^a \beta^b + \beta^b \beta^a&=& 2 \eta^{ab} e.\nonumber
\end{eqnarray}
Then both sets (\ref{gmatr3}) generate bases of Clifford algebra and $\gamma^{1\ldots n}$, $\beta^{1\ldots n}$ equal $\pm e^{1\ldots n}$.

Moreover, there exists a unique (up to multiplication by an invertible element of the center of the Clifford algebra) invertible element $T\in \cl^\F_{\Even}(p,q)$
(and also another $T\in \cl^\F_{\Odd}(p,q)$ which is a product of the previous element and $e^{1\ldots n}$) such that
\begin{itemize}
  \item \begin{eqnarray}
\gamma^{a}=T^{-1}\beta^a T,\qquad \forall a=1, \ldots, n\label{pauli2}
\end{eqnarray}
if and only if $\beta^{1\ldots n}=\gamma^{1\ldots n}$,
  \item \begin{eqnarray}
\gamma^{a}=-T^{-1}\beta^a T,\qquad \forall a=1, \ldots, n \label{pauli22}
\end{eqnarray}
if and only if $\beta^{1\ldots n}=-\gamma^{1\ldots n}$.
\end{itemize}

Moreover, in both cases such $T$ can be written in the following form
$$T=\sum_{A\in\I_{\Even}}\beta^A F \gamma_A,$$
where $F$ is such element from the set
$\{\gamma^A,\quad A\in\I_{\Even}\}$ (or also such element from the set $\{\gamma^A,\quad A\in\I_{\Odd}\}$),
that $\sum_{A\in\I_{\Even}}\beta^A F \gamma_A\neq 0$.
\end{thm}

\proof We have only 2 cases $\gamma^{1\ldots n}=\pm e^{1\ldots n}$ (not $\gamma^{1\ldots n}=\pm e$ and $\pm ie$) because $\gamma^a\in\cl^\F_{\Odd}(p,q)$, $n$ - odd. Then we use Theorems \ref{theoremPauliOddReal} and \ref{theoremPauliOddCompl}.
$\blacksquare$

Now we are interested in the other relation between 2 sets of elements: $\gamma^{a}=T^{\wedge -1}\beta^a T$, where $\wedge$ is a grade involution (see (\ref{wedge})). Let formulate theorems for this case.

\begin{thm}\label{theoremPauliEvOdd2} Consider real (or, respectively complex) Clifford algebra $\cl^\F(p,q)$ of even dimension $n=p+q$. Let two sets of odd Clifford algebra elements
\begin{eqnarray}
\gamma^a,\, \beta^a \in \cl^\F_{\Odd}(p,q),\qquad a=1, 2, \ldots, n\label{gmatr2}
\end{eqnarray}
satisfy conditions
\begin{eqnarray}
\gamma^a \gamma^b + \gamma^b \gamma^a&=& 2 \eta^{ab} e, \label{gsootn27}\\
\beta^a \beta^b + \beta^b \beta^a&=& 2 \eta^{ab} e.\nonumber
\end{eqnarray}

Then both sets generate bases of Clifford algebra and $\gamma^{1\ldots n}$, $\beta^{1\ldots n}$ equal $\pm e^{1\ldots n}$.

There exists a unique (up to  multiplication by a nonzero real (or, respectively complex) constant) invertible element $T\in\cl^\F(p,q)$ such that
\begin{eqnarray}
\gamma^{a}=T^{\wedge -1}\beta^a T,\qquad \forall a=1, \ldots, n,\label{pauli27}
\end{eqnarray}
where $\wedge$ is a grade involution. 

Moreover,
\begin{itemize}
          \item $T\in\cl^\F_{\Even}(p,q)$ if $\beta^{1\ldots n}=\gamma^{1\ldots n}$,
          \item $T\in\cl^\F_{\Odd}(p,q)$ if $\beta^{1\ldots n}=-\gamma^{1\ldots n}$.
\end{itemize}
Moreover, $T$ can be written in the following form
\begin{itemize}
  \item $T=\beta^A F \gamma_A\quad$ if $\beta^{1\ldots n}=\gamma^{1\ldots n}$,
  \item $T=(-1)^{|A|}\beta^A F \gamma_A\quad$ if $\beta^{1\ldots n}=-\gamma^{1\ldots n}$,
\end{itemize}
where $F$ is such element from the set
\begin{itemize}
  \item $\{ \gamma^A, A\in\I_{\Even}\}\quad$ if $\beta^{1\ldots n}=\gamma^{1\ldots n}$,
  \item $\{ \gamma^A, A\in\I_{\Odd}\}\quad$ if $\beta^{1\ldots n}=-\gamma^{1\ldots n}$,
\end{itemize}
 that $T$ constructed on its $F$ is nonzero.
\end{thm}

\proof 1) If $\beta^{1\ldots n}=\gamma^{1\ldots n}$, using Theorem \ref{theoremPauliEvOdd}, $\exists T^\wedge=T\in\cl^\F_{\Even}(p,q)$ such that $\gamma^a=(T^{\wedge})^{-1}\beta^a T$.

2) If $\beta^{1\ldots n}=-\gamma^{1\ldots n}$, then $\exists T^\wedge=-T\in\cl^\F_{\Odd}(p,q)$ such that $\gamma^a=-T^{-1}\beta^a T=(T^{\wedge})^{-1}\beta^a T$.
So, we obtain Theorem \ref{theoremPauliEvOdd2}. $\blacksquare$

\begin{thm}\label{theoremPauliOddOdd2} Consider real (or, respectively complex) Clifford algebra $\cl^\F(p,q)$ of odd dimension $n=p+q$. Let two sets of odd Clifford algebra elements
\begin{eqnarray}\gamma^a,\, \beta^a\in\cl^\F_{\Odd}(p,q),\qquad a=1, 2, \ldots, n.\end{eqnarray}
satisfy conditions
\begin{eqnarray}
\gamma^a \gamma^b + \gamma^b \gamma^a&=& 2 \eta^{ab} e, \\
\beta^a \beta^b + \beta^b \beta^a&=& 2 \eta^{ab} e.\nonumber
\end{eqnarray}

Then both sets generate bases of Clifford algebra and $\gamma^{1\ldots n}$, $\beta^{1\ldots n}$ equal $\pm e^{1\ldots n}$.

There exists a unique (up to multiplication by a nonzero real (or, respectively complex) constant) invertible element $T\in\cl^\F(p,q)$ such that
\begin{eqnarray}
\gamma^{a}=(T^\wedge)^{-1}\beta^a T,\qquad \forall a=1, \ldots, n,\label{pauli444}
\end{eqnarray}
where $\wedge$ is a grade involution. 

Moreover,
\begin{itemize}
  \item $T\in \cl^\F_{\Even}(p,q)$ if $\beta^{1\ldots n}=\gamma^{1\ldots n}$,
  \item $T\in \cl^\F_{\Odd}(p,q)$ if $\beta^{1\ldots n}=-\gamma^{1\ldots n}$.
\end{itemize}

Moreover, $T$ can be written in the following form
$$T=\sum_{A\in\I_{\Even}}\beta^A F \gamma_A,$$
where $F$ is such element from the set
\begin{itemize}
  \item $\{\gamma^A,\quad A\in\I_{\Even}\}$ if $\beta^{1\ldots n}=\gamma^{1\ldots n}$,
  \item $\{\gamma^A,\quad A\in\I_{\Odd}\}$ if $\beta^{1\ldots n}=-\gamma^{1\ldots n}$,
\end{itemize}
that $\sum_{A\in\I_{\Even}}\beta^A F \gamma_A\neq 0$.
\end{thm}

\proof 1) If $\beta^{1\ldots n}=\gamma^{1\ldots n}$, using Theorem \ref{theoremPauliOddOdd}, $\exists T^\wedge=T\in\cl^\F_{\Even}$ such that $\gamma^a=(T^\wedge)^{-1}\beta^a T$.

2) If $\beta^{1\ldots n}=-\gamma^{1\ldots n}$, then $\exists T^\wedge=-T\in\cl^\F_{\Odd}$ such that $\gamma^a=-T^{-1}\beta^a T=(T^\wedge)^{-1}\beta^a T$.
So, we obtain Theorem \ref{theoremPauliOddOdd2}. $\blacksquare$

Note that in Theorem \ref{theoremPauliOddOdd2} we have unique element $T$ up to multiplication by a constant (not unique up to multiplication by an invertible element of the center of the Clifford algebra as in Theorems \ref{theoremPauliOddReal} and \ref{theoremPauliOddCompl}).

\section{Orthogonal and spinor groups}

Consider pseudo-orthogonal group and its subgroup - special pseudo-orthogonal group
\begin{eqnarray}
\O(p,q)&=&\{A\in \Mat(n,\R)\,|\, A^T \eta A=\eta\}\nonumber\\
\SO(p,q)&=&\{A\in \Mat(n,\R)\,|\, A^T \eta A=\eta,\, \det A=1\}\nonumber
\end{eqnarray}
where $p$ and $q$ - nonnegative integer numbers such that $p+q=n,\ n\geq 1$ and $\eta=||\eta^{ab}||=\diag(1,\ldots,1,-1,\ldots,-1)$ is a diagonal matrix, whose diagonal contains $p$ elements equal to $+1$ and $q$ elements equal to $-1$.

We will consider signatures $(p,q)$ of Euclidian space $V$ where the first $p$ coordinates are time coordinates and the last $q$ coordinates are space coordinates.

Consider determinant of matrix consisting of the elements standing on the intersections of rows $k_1, \ldots k_i$ and columns $l_1, \ldots l_j$ . We called it minor and denote it by $A^{k_1 \ldots k_i}_{l_1 \ldots l_i}$.

Consider {\it orthochronous}, {\it orthochorous} and {\it special orthochronous} groups (see, for example, \cite{Benn:Tucker})
\begin{eqnarray}
\O_{\uparrow}(p,q)&=&\{A\in \O(p,q)\,|\, A^{1\ldots p}_{1\ldots p}> 0\},\nonumber\\
\O_{\downarrow}(p,q)&=&\{A\in \O(p,q)\,|\, A^{p+1\ldots n}_{p+1\ldots n}> 0\},\nonumber\\
\SO_{\uparrow\downarrow}(p,q)&=&\{A\in \O(p,q)\,|\, A^{1\ldots p}_{1\ldots p}> 0,\, \det A=1\}.\nonumber
\end{eqnarray}

Note, that $\SO_{\uparrow\downarrow}(p,q)=\SO_+(p,q)$ is the connected component of the identity. Orthochronous group consist of transformations that preserve time orientation, orthochorous group consist of transformations that preserve parity. Note, that if elements of $\SO(p,q)$ are orthochronous then they must be parity preserving (orthochorous).

It can be shown that pseudo-orthogonal group $\O(p,q)$ consists of 4 connected components for $p, q\neq 0$ (see \cite{Benn:Tucker}):
$$\O(p,q)=\SO_{\uparrow\downarrow}(p,q)\sqcup\O_{\uparrow}'(p,q)\sqcup\O_{\downarrow}'(p,q)\sqcup\SO'(p,q),$$
where
\begin{eqnarray}
\O_{\uparrow}'(p,q)&=&\O_{\uparrow}(p,q)\setminus\SO_{\uparrow\downarrow}(p,q),\nonumber\\
\O_{\downarrow}'(p,q)&=&\O_{\downarrow}(p,q)\setminus\SO_{\uparrow\downarrow}(p,q),\nonumber\\
\SO'(p,q)&=&\SO(p,q)\setminus\SO_{\uparrow\downarrow}(p,q).\nonumber
\end{eqnarray}

Consider homomorphism $ad: \cl^{\R\times}(p,q)\rightarrow \operatorname{End}\cl^\R(p,q)$ acting on the group of invertible Clifford algebra elements in the following way
$s\mapsto ad_s$ where
$$ad_s x=sxs^{-1},\qquad x\in\cl^\R(p,q).$$

Also consider homomorphism $\st{\wedge}{ad}: \cl^{\R\times}(p,q)\rightarrow \operatorname{End}\cl^\R(p,q)$ acting on the group of invertible Clifford algebra elements in the following way
$s\mapsto \st{\wedge}{ad}_s$ where $$\st{\wedge}{ad}_s x=s^{\wedge}xs^{-1},\qquad x\in\cl^\R(p,q).$$

Denote by $\cl^{\R\times}_{\Even}(p,q)$ and $\cl^{\R\times}_{\Odd}(p,q)$ sets of even and odd invertible Clifford algebra elements.
Consider {\it Lipschitz group}
$$\Gamma^{\pm}(p,q)=\{s\in\cl^{\R\times}_{\Even}(p,q)\cup\cl^{\R\times}_{\Odd}(p,q)| \forall x\in\cl^\R_1(p,q), sxs^{-1}\in\cl^\R_1(p,q)\}$$
and its special subgroup
$$\Gamma^{+}(p,q)=\{s\in\cl^{\R\times}_{\Even}(p,q)| \forall x\in\cl^\R_1(p,q), sxs^{-1}\in\cl^\R_1(p,q)\},$$
where $\cl^\R_1(p,q)$ is subspace of Clifford algebra elements of rank 1 (see (\ref{U}) and (\ref{bigop})).

We have the following 5 different {\it spinor groups}
\begin{eqnarray}
\Pin(p,q)&=&\{T\in\Gamma^{\pm}| T^\sim T=\pm e\}=\{T\in\Gamma^{\pm}| T^{\sim\wedge} T=\pm e\}, \nonumber\\
\Pin_{\downarrow}(p,q)&=&\{T\in\Gamma^{\pm}| T^\sim T=+e\},\nonumber\\
\Pin_{\uparrow}(p,q)&=&\{T\in\Gamma^{\pm}| T^{\sim\wedge} T=+e\},\label{spin}\\
\Spin(p,q)&=&\{T\in\Gamma^{+}| T^\sim T=\pm e\}=\{T\in\Gamma^{+}| T^{\sim\wedge} T=\pm e\},\nonumber\\
\Spin_{\uparrow\downarrow}(p,q)&=&\{T\in\Gamma^{+}| T^\sim T=+e\}=\{T\in\Gamma^{+}| T^{\sim\wedge} T=+e\}.\nonumber
\end{eqnarray}

The group $\Pin(p,q)$ consists of 4 components for $p, q\neq 0$:
$$\Pin(p,q)=\Spin_{\uparrow\downarrow}(p,q)\sqcup\Pin_{\uparrow}'(p,q)\sqcup\Pin_{\downarrow}'(p,q)\sqcup\Spin'(p,q),$$
where
\begin{eqnarray}
\Pin_{\uparrow}'(p,q)&=&\Pin_{\uparrow}(p,q)\setminus\Spin_{\uparrow\downarrow}(p,q),\nonumber\\
\Pin_{\downarrow}'(p,q)&=&\Pin_{\downarrow}(p,q)\setminus\Spin_{\uparrow\downarrow}(p,q),\nonumber\\
\Spin'(p,q)&=&\Spin(p,q)\setminus\Spin_{\uparrow\downarrow}(p,q).\nonumber
\end{eqnarray}

The following well-known statement shows the relation between spinor and orthogonal groups.

\begin{thm} The homomorphism $\st{\wedge}{ad}$ acting
from
\begin{eqnarray}
\O(p,q),\, \SO(p,q),\, \SO_{\uparrow\downarrow}(p,q),\, \O_{\uparrow}(p,q),\, \O_{\downarrow}(p,q)\label{ort}
\end{eqnarray}
to (respectively)
\begin{eqnarray}
\Pin(p,q),\, \Spin(p,q),\, \Spin_{\uparrow\downarrow}(p,q),\, \Pin_{\uparrow}(p,q),\, \Pin_{\downarrow}(p,q)\label{spinor}
\end{eqnarray}
is surjective with kernel $\{\pm 1\}$. Moreover, spinor groups are double covers of corresponding orthogonal groups.

\end{thm}

We can use the formula
\begin{eqnarray}
T^{\wedge}e^a T^{-1}=p^a_be^b,\label{svyaz}
\end{eqnarray}
that associates a pair of elements $\pm T$ of spinor group (\ref{spinor})
to a single matrix $P=||p^a_b||$ of corresponding orthogonal group (\ref{ort}).

\section{Calculation of elements of spinor groups}

Using Theorems \ref{theoremPauliEvOdd2} and \ref{theoremPauliOddOdd2} we can obtain algorithm for computing elements of spinor groups that correspond to given elements of orthogonal groups.
We can use formula (\ref{svyaz}) that associates a pair of elements $\pm T$ of corresponding spinor group (\ref{spinor})
to a single matrix $P=||p^a_b||$ of corresponding orthogonal group (\ref{ort}).

\begin{thm}\label{theoremSpinn} Consider real Clifford algebra $\cl^\R(p,q)$ of dimension $n=p+q$. Let $P\in\O(p,q)$ be an orthogonal matrix. Then we can find elements $\pm T\in\Pin(p,q)$ such that $\st{\wedge}{ad}(\pm T)=P$ in the following way.

Consider the following set of Clifford algebra elements
$$\beta^a=p^a_be^b,\qquad P=||p^a_b||.$$

1) In the case of even $n$ we can always find $T\in\Gamma^{\pm}$ among elements
\begin{itemize}
  \item $T=\beta^A F e_A\quad$ if $\beta^{1\ldots n}=e^{1\ldots n}$,
  \item $T=(-1)^{|A|}\beta^A F e_A\quad$ if $\beta^{1\ldots n}=-e^{1\ldots n}$,
\end{itemize}
where $F$ is such element from the set
\begin{itemize}
  \item $\{ e^A, A\in\I_{\Even}\}\quad$ if $\beta^{1\ldots n}=e^{1\ldots n}$,
  \item $\{ e^A, A\in\I_{\Odd}\}\quad$ if $\beta^{1\ldots n}=-e^{1\ldots n}$,
\end{itemize}
that $T$ constructed on its $F$ is nonzero.

Taking into account conditions $T^{\sim} T=\pm e$ (or $T^{\sim\wedge} T=\pm e$) we can find two elements $\pm T$ from the group $\Pin(p,q)$ that correspond to the orthogonal matrix $P$.

2) In the case of odd $n$ we act the same way. We find element $T$ among elements
$$T=\sum_{A\in\I_{\Even}}\beta^A F e_A,$$
where $F$ is such element from the set
\begin{itemize}
  \item $\{e^A,\quad A\in\I_{\Even}\},$ if $\beta^{1\ldots n}=e^{1\ldots n}$,
  \item $\{e^A,\quad A\in\I_{\Odd}\},$ if $\beta^{1\ldots n}=-e^{1\ldots n}$,
\end{itemize}
that $\sum_{A\in\I_{\Even}}\beta^A F e_A\neq 0$.

Further we take into account conditions $T^{\sim} T=\pm e$ or $T^{\sim\wedge} T=\pm e$.
\end{thm}

\proof  The set $\beta^a=p^a_be^b$, $P\in\O(p,q)$, satisfy conditions $\beta^a\beta^b+\beta^b\beta^a=2\eta^{ab}e$. So, we can use Theorems \ref{theoremPauliEvOdd2} and \ref{theoremPauliOddOdd2}. Expressions $T^{\sim} T$ and $T^{\sim\wedge} T$ are elements of the rank $0$ for this element $T$ (see, for example, \cite{Benn:Tucker}). $\blacksquare$

Note, that we can calculate elements of all subgroups of group $\Pin(p,q)$:
$$\Spin(p,q),\, \Pin_{\uparrow}(p,q),\, \Pin_{\downarrow}(p,q),\, \Spin_{\uparrow\downarrow}(p,q)$$
in the same way. So, now we can always solve nonlinear equation (\ref{svyaz}) in Clifford algebra with respect to $T$ and find elements of spinor groups.

Let's give some examples.

Consider Clifford algebra $\cl(p,q)$ with arbitrary odd $p$ and odd $q$, $p+q=n$. We have time-reversal matrix
$$T=-\eta=-\diag(1, \ldots ,1, -1, \ldots, -1)\in\O'_{\downarrow}(p,q),$$
parity-reversal matrix
$$P=\eta\in\O'_{\uparrow}(p,q)$$ and $$T*P=-\eta* \eta={\bf -1}\in\SO'(p,q).$$
Then using algorithm from this section we can obtain the following:
\begin{eqnarray}
\st{\wedge}{ad}(\pm e^{1\ldots p})&=&-\eta,\nonumber\\
\st{\wedge}{ad}(\pm e^{p+1\ldots n})&=&\eta,\nonumber\\
\st{\wedge}{ad}(\pm e^{1\ldots n})&=&-\eta*\eta={\bf -1},\nonumber
\end{eqnarray}
where
\begin{eqnarray}
\pm e^{1\ldots p}&\in&\Pin'_{\downarrow}(p,q),\nonumber\\
\pm e^{p+1\ldots n}&\in&\Pin'_{\uparrow}(p,q),\nonumber\\
\pm e^{1\ldots n}&\in&\Spin'(p,q).\nonumber
\end{eqnarray}

Spin groups and Clifford algebras are used in different branches of modern mathematics and physics: field theory, robotics, signal processing, computer vision, chemistry, celestial mechanics, electrodynamics, etc. The author hopes that the proposed algorithm for calculating the elements of spinor groups will be used in different applications.

Also generalized Pauli's theorem can be useful in different applications of mathematical physics. For example, when we consider Weyl, Majorana and Majorana-Weyl spinors in arbitrary dimension $n=p+q$. Using generalized Pauli's theorem we can give analogues of Dirac, Majorana and charge-conjugation in the case of arbitrary space dimension and signature.


\subsection*{Acknowledgment}
The author is grateful to N.G.Marchuk for the constant attention to this work.

\end{document}